\documentstyle[preprint,aps]{revtex}
\begin{document}
\draft
\title {
Hole Dynamics in a Quantum Antiferromagnet\\
beyond the Retraceable Path Approximation}
\author{  Q. F. Zhong and S. Sorella}
\address{
International School for Advanced Study\\
Via Beirut 4, 34013 Trieste, Italy}
\author{A. Parola}
\address{Dipartimento di Fisica, Universit\'a di Milano, Via Celoria 16,
Milano, Italy.}
\date{SISSA preprint: 156/93/CM/MB, BABBAGE: cond-mat/9310023}
\maketitle
\begin{abstract}
The one-hole spectral weight for two chains and two dimensional lattices
is studied numerically using a new  method of analysis of the spectral
 function
within the Lanczos iteration scheme: the Lanczos spectra decoding method.
This technique is applied to
the $t-J_z$ model for $J_z \to 0$, directly in the  infinite size lattice.
By a careful investigation of the first 13 Lanczos steps and the first 26
 ones
for the two dimensional and the two chain  cases respectively,
 we get several new features of the one-hole spectral weight.
A sharp incoherent peak with a clear momentum dispersion is identified,
together with a second broad  peak at higher energy. The spectral weight is
finite up to the Nagaoka energy where it vanishes in a non-analytic way.
Thus the lowest energy of
one hole in a quantum antiferromagnet is degenerate with
the Nagaoka energy in the thermodynamic limit.
\end{abstract}
\pacs{75.10.Jm,75.40.Mg,71.10.+x}
\narrowtext

After the discovery of superconductivity in
materials which exhibit antiferromagnetic (AF) long range order in the
insulating phase, much attention
has been given to the study of strongly correlated  system at low
doping. Despite the enormous amount of work, the dynamical
properties of one hole in a quantum antiferromagnet (QAF)
are still a subject of debate \cite{sorella,poilblanc,dagotto}.

A satisfactory model of a QAF is provided by the Heisenberg hamiltonian
which, in 2D, is known to show N\'eel long range order. In the following
we mimic the properties of a QAF by use of the simpler AF Ising hamiltonian
whose {\sl exact} ground state is the N\'eel state and we address the
problem of the propagation of a hole in the $J_z\to 0$ limit of the
$t-J_z$ model defined by:
\begin{equation}
 H=-t\sum_{<i,j>,\sigma} ( c_{i \sigma}^\dagger c_{j \sigma}\,+ \,h.c.)
+J_z \sum_{<i,j>}
{ S_i^z \, S_j^z}
\label{tjz}
\end{equation}
where the constraint of no double occupancy is understood.
Both the two chain (2C) and the two dimensional square lattice (2D)
will be considered. The Nagaoka (NK) theorem \cite{nagaoka},
provides a lower bound $e_F$ to the
ground state energy of one hole in the $J_z\to 0$ limit:
$e_F=-z t$, where $z$ is the coordination number of the lattice
($z=3$ for the 2C problem and $z=4$ for the 2D case).

A complete description of the one particle spectrum,
not limited  only to the ground state,
was first given in the seminal work of Brinkman and Rice (BR),
where the so called ``retraceable path approximation"(RP) was
introduced\cite{br}.
In this approximation the spectral weight
is completely incoherent and the hole is essentially localized
without  dispersion.
The RP is exact in 1D ( where NK  theorem does not apply)
and recently it has also been shown to
be exact  in the limit of infinite spatial
dimensionality\cite{metzner,vollhardt}.
However, no analytical  solution is available for finite $z>2$
and  it is not clear how accurate the RP is in these cases.
Indeed, in finite dimensions $D>1$ several one hole paths exist,
allowing for the propagation of the hole \cite{trugman}.

Recently, the problem of hole propagation has been addressed by use of
the quite efficient Lanczos technique in 2D lattices with up to 26 sites
\cite{poilblanc} or with a promising approximate technique up to 50 sites
\cite{riera}.
 Despite the remarkable numerical effort, still the number of
sites appears to be too small for a precise determination
of the properties of the spectral function or for the ground state
properties for $J_z \to 0$.

In this work, we  present a numerical study of the one hole spectral weight
by means of
the Lanczos iteration scheme, applied directly to the infinite system
\cite{trugman}.
 We introduce  a new method for the calculation of the spectral weight, that we
name the Lanczos spectra decoding method
(LSD). This new method  combines the power of the Lanczos algorithm with
some general properties of the spectral function.

As well known, the Lanczos technique consists of a partial diagonalization of
the
hamiltonian in the basis generated by the vectors ${\bf s_n}=H^n \vert \Psi_T
\rangle $ for $n=0,1,...,N$, where
 $\vert \Psi_T \rangle$ is a trial state.
After orthogonalization of the vectors $\bf{s_n}$, the hamiltonian reduces to
tridiagonal form and the resulting orthogonal basis $\{{\bf e_i} \}$
can be iteratively  calculated\cite{baroni}.
In this restricted basis it is  straightforward to compute
static and dynamical correlation functions. Usually,
convergence is reached for N much smaller than the dimension of the
Hilbert space, justifying the success of the method.
In an infinite system this procedure can be applied as well,
provided  the Hilbert space generated by the iterative application of $H$ to
the trial state  remains finite. This is in fact the case for the ``simple''
$t-J_z$ model if we consider the basis of states where the hole is located at
the origin and the z-component of the spin is defined in any other site.
The application of the hamiltonian to one of these states
generates at most $z$ new elements of the basis and,
after $N$ steps, the Hilbert space is finite having at most dimension $z^N$.
However, this exponential growth makes the problem intractable even
for relatively small $N$ . Fortunately many of the
generated states appear several times during the expansion process of
 the Hilbert space which in fact turns out to be considerably smaller
than the previous estimate. Another reduction factor can be gained
 by implementing the translation  symmetry of the hamiltonian.
Translation operators can be always used to move the hole at the
origin of the lattice and only the corresponding basis elements must
be stored.

By applying the above strategy,  the dimension of the Hilbert space grows
much slower than $z^N$ and in fact we are able to go
up to $N=26$  for the 2C case and $N=13$ for the 2D case with an Hilbert space
dimension at most equal to $\sim 12.2 \times 10^6$.
The 2C model is much easier to study with this method, though the
properties of the spectral weight are rather similar to the 2D case as we will
show in the following.
Further details of this calculation will be published elsewhere\cite{noi}.

We apply the Lanczos algorithm to the $t-J_z$ model, using as
trial state $|\psi_T>$ the exact ground state of the undoped system
with a hole of definite momentum $k$, i.e.
$\vert \Psi_T \rangle={1\over \sqrt{L/2}} \sum_{R\in A}
 e^{i {\bf k \cdot R}}  c_{R \uparrow}
\vert \Phi \rangle$
where $\vert \Phi \rangle $ is the N\'eel state with  spin up electrons
in the $A$ sublattice and $L$ is the size of the lattice, where
periodic boundary conditions are assumed.

The spectral weight can be formally calculated at fixed $N$ and reads:
\begin{eqnarray}
A(k,\omega)&=&{\rm Im} {{1} \over {\pi}} \langle \Psi_T
\vert {{1} \over {\omega -H-i \delta }} \vert \Psi_T \rangle \nonumber \\
 &=& \sum_{i=0}^N \vert \langle \Psi_i
\vert \Psi_T \rangle \vert ^2 \delta ( \omega - E_i) \label{sweight}
 \end{eqnarray}
where $E_i$ and $\vert \Psi_i \rangle$ are eigenvalues and eigenstates
of $H$ in the restricted Lanczos basis.
We expect that for $J_z\to 0$ the spectral function is
completely incoherent in the infinite size limit and we know that
by the NK  theorem  $A(k,\omega)$ is identically zero outside the interval
$e_F < \omega < -e_F$. As a result  of the finiteness of our
Hilbert space, at fixed $N$ we get a sum of
$\delta$-functions in the spectral weight. This feature also appears in the
exact spectral function of finite systems and, in that framework, an
estimate of the thermodynamic limit is obtained by
smoothing the $\delta$-functions in Eq.~(\ref{sweight}) with
lorenzians of a given small width\cite{poilblanc,dagotto}:
$\delta(\omega-E_i)\,\, \to  \,\, {\rm Im} {{\pi^{-1}} \over
{\omega-E_i-i\delta}}$.
For small  $\delta$  reasonable results can be obtained, provided  the
resolution of the energy levels becomes  much smaller than $\delta$ for
large $N$.
In our case, however, we cannot reach very large $N$ and a more efficient
method for evaluating the  spectral weight is necessary. The $N\to\infty$ limit
of the spectral function may contain a coherent part only if in this limit some
$Z_i= \vert\langle \Psi_i \vert \Psi_T \rangle \vert ^2 $ attains
a finite value. We have verified that for both the $2C$ and the square lattice,
all $Z_i$ tend to zero and therefore the spectral function is
completely incoherent as expected.
The spectral weight can be always written as a product of two functions:
\begin{equation}
A(k,\omega)=Z(\omega)\rho_L(\omega) \label{zrho}
\end{equation}
where
\begin{equation}
\rho_L(\omega)={{1}\over {N+1}} \sum_{i = 0}^N
\delta (\omega-E_i) \label{rho}
\end{equation}
and $Z(\omega)$ is defined at discrete points:
\begin{equation}
Z(\omega)=(N+1)\vert \langle \Psi_i \vert \Psi_T \rangle \vert^2
\label{zeta}
\end{equation}
for $\omega=E_i$.
If $A(k,\omega)$ is incoherent and
the Lanczos eigenvalues are set in ascending order $E_{i+1}>
E_i$, $ Z(\omega)$ describes a well behaved function of
$\omega$ as well as the coarse grained
Lanczos density of states (LDOS) which can be estimated from Eq. (4) as
\begin{equation}
\displaystyle \rho_L ( \bar \epsilon_i =
{{E_i+E_{i+1}} \over {2}}) \,=\,{{1} \over{(N+1)(E_{i+1}-E_i)}}.
\label{rhod}
\end{equation}
Analogously, $Z(\omega)$
can be then interpolated linearly at energies $\bar \epsilon_i$:
\begin{equation}
Z(\bar \epsilon_i) = (Z_{i+1}+Z_i)/2.
\label{zetad}
\end{equation}
Eqs.~ (\ref{rhod},\ref{zetad}) are accurate to  $O(1/N^2)$ and
better fits  are possible using more data  for the interpolation
procedure.
$A(k,\omega)$  easily follows  from Eq.~(\ref{zrho}).
At the end, we can verify the sum rule $\int A(k,\omega)d \omega=1  $,
as a check  for the accuracy of the calculation.
This is all about  the LSD method.

In order to test  LSD  we have applied it to an exactly solvable problem.
If we neglect all closed paths in the averages of $<\psi_T |H^n | \psi_T>$
 the resulting Green's function
is the known BR one. On the other hand, this  is the {\em exact}
Green's function in the Bethe lattice of coordination $z$.
LSD  is perfectly defined in this case.
In fact, it is possible to  compute analytically $Z$ and $\rho_L$
in Eqs.~(\ref{rho},\ref{zeta}) due to the simple structure of
the Lanczos matrix in the Bethe lattice case:
$\langle {\bf e_i} \vert H \vert {\bf e_{i+1}}\rangle= \sqrt {z-1}$
for $i > 0$ and $\langle {\bf e_0} \vert H \vert {\bf e_1}\rangle=\sqrt{z}$
are the only non-zero  upper diagonal matrix elements in the orthogonal
Lanczos basis $\{ {\bf e_i} \} $.

In Fig.~\ref{fig1} we show a comparison of the spectral weight obtained
by smearing out the $\delta$-functions--usual method--
 and by applying our scheme.
We see that the agreement of our data with the exact results is very good and
we easily resolve the first peak in $A(\omega)$ (dispersionless in this case)
for the 2C case.
By contrast the usual method cannot give accurate  results with the
available number $N$ of Lanczos steps.

In the following, we present the results for the 2C and the 2D case, obtained
using this new method to evaluate the spectral weight, without any
approximation. We show in Fig.~\ref{fig2}
the spectral  weight for $k=(0,0)$ obtained with different
numbers of Lanczos steps. For the 2C case, our results are well converged
and we find a sharp peak  located at an energy close to the BR one
$e_{BR}=-2 t \sqrt{ z-1} $
and, surprisingly, a second peak at energy $\sim -t$. In the 2D case, the
spectral weight looks similar, although the first peak is rather small.

In 1D, the exact BR solution leads only to one peak but with a divergent
 spectral weight $\sim {1 \over \sqrt {\omega_c-\omega}} $ at the bottom
 $\omega_c$ of the band.
Already in the 2C case such a divergence disappears within the RP, as well as
in our numerical scheme, which includes all closed-loop paths.
 In fact the peak in
Fig.~\ref{fig2}  does not depend  much on the number $N$  of Lanczos steps.

In  Fig. ~\ref{fig3}  we see that
the first peak in the spectral function has a remarkable dispersive
feature although the bottom of the spectrum appears  $k-$independent.
The dispersion of the first  peak  is not present neither in 1D or in
infinite
dimension\cite{metzner,vollhardt}
and the importance to go beyond the RP  is clear even in 2D.

The one particle density of states (DOS) can be computed
either  by integrating  $A(k,\omega)$ over $k$
or equivalently by a direct evaluation of ${\rm Im} G(R=0,\omega). $
As it is shown in Fig.2 for the 2C case, our results present some small
deviation to the DOS in RP approximation especially when large closed paths
are allowed.
In 2D however the BR solution seems already quite accurate,
at least away from the band tails.

Another open problem is whether the band edge of a hole in a QAF coincides with
the NK  energy and  how the spectral weight
vanishes at the band edges.
At finite $N$ the lowest eigenvalue of the hamiltonian $E_N$ restricted to the
Lanczos basis is of course a variational upper bound to
the lowest eigenstate  non-orthogonal to $|\psi_T>$.
Note that the NK  state is orthogonal to $|\psi_T>$ only in the infinite
system and $E_{\infty}=\lim\limits_{N\to \infty} E_N$  may be different from
$e_F$, contrary to the finite
size case. Thus $E_\infty$ is a definition of the one hole energy in a quantum
antiferromagnet and since it coincides with the smallest energy $\omega$
where $A(k,\omega)$ vanishes.
In order to have a good estimate of this energy $E_\infty$  it is useful
to know what are the leading corrections of the quantity
$\Delta _N = E_N -  E_{\infty}$ for $N\to \infty$.
The way $\Delta _N$ vanishes for $N \to \infty$ is related to the form of
the LDOS  at low energy. In the BR case the exact solution gives
$\rho_L(\epsilon) \sim \epsilon^{-1/2}$ (see Fig.2).
Thus, using Eqs.~(\ref{rhod}), $\Delta_N ^{-1/2} \sim { 1\over N \Delta _N} $,
 yielding
$\Delta _N \sim {1\over N^2} $. In the general case we have numerical
evidence of a finite LDOS, and the same argument
determines $\Delta _N \sim { 1\over N}.$

We have plotted in Fig.4(a) the estimated ground state energies
 as a function of $1/N$ for several momenta  for the 2C case.
 Many of the estimated Lanczos energies -- exact upper bound of the
true ground state energy-- are clearly below the BR energy (even for the 2D
case not shown in the figure\cite{noi}).
Thus,  a previous suggestion that the one hole
energy in a QAF should be  close to $e_{BR}$  \cite{hasegawa,emery}
{\em is not confirmed}  by our numerical results.
In Fig.4(a) it is a remarkable property that all the extrapolated energies
 are very close to the NK  energy, independent of the momentum of the
hole, $E_{\infty} = -3 \pm 0.02$ although the spin configuration is
antiferromagnetic. This clearly suggests that the spectral  weight
is finite  up to the NK  energy. Moreover the fact that the corresponding
quasiparticle weight becomes very small close to the NK  energy (e.g.
$Z =0.02$ for $E_{26}=-2.92$ and $p=0$) may indicate that the vanishing of
 the spectral weight is non-analytic.
Indeed a least square fit of our data for the band tail is more consistent
with an exponential  vanishing rather  than  with a  power law.

In order to check the estimated
$J_z \to 0$ energy we have also computed $E_N$ for the finite $J_z$ model,
where due to the localization of the hole in the linear ``string''
potential\cite{vollhardt} $E_N$ converges exponentially
with $N$ and $N=26$ gives already a
very accurate estimate (within 0.1 \%) for $J_z >0.1$. For smaller $J_z$
( $ 0.02 \le  J_z < 0.1$ )
 the same accuracy is obtained by fitting the Lanczos data with
$\Delta_N \propto {e^{- \,{\it const.}\,\, N}\over N } $, which interpolates
consistently the $\Delta_N \to 0$ convergence in the limit $J_z=0$ and in the
finite $J_z$  case.
In the string picture the one hole energy for small $J_z$ should be
$E=a +b {J_z\over t}^{\theta}$ with $a=e_{BR}$  and $\theta=2/3$
independent of $z$.  Our data shown in
Fig.~4(b)  are clearly consistent with the BR exponent $2/3$, although
$a=e_F$. The next leading corrections to the contribution $J_z^{2/3}$,
are also very important. For small $J_z$ in the 2C case
we do not find any transition to a
phase separated (PS) polaron state\cite{emery}:
 $E_{PS}=-3t +c ({J_z\over t})^{2/3}$, with $c/t= 3 (\pi/2)^{2/3}=4.05 $.
Previous numerical work\cite{barnes} is in qualitative agreement with our
results although the small size studied\cite{barnes}  was not enough to detect
the  $J_z\to 0$  {\em smooth} crossover to the NK  energy, that we found in
the 2C case.

In summary,
we have presented here a successful attempt to go beyond the RP for the hole
dynamics in a  QAF. A new Lanczos-type
of analysis of the spectral weight  enabled us to get very accurate results for
the 2C problem and qualitatively similar ones for the 2D case.
A clear dispersion of  the main incoherent peak of
$A(k,\omega)$ both for the 2C and the 2D case was found.
The variational argument about phase separation
on the $t-J_z$ model (or $t-J$)
for small $J_z$ was based on the assumption that the one hole
energy in a QAF is smaller than the NK  energy $e_F$
by a finite amount\cite{emery}.
Instead our results strongly suggest that $E_{\infty} = e_F$ for $J_z \to 0$.
In fact as it is shown in Fig.4(b) the one hole energy in a phase separated
polaron is always well above the exact estimated ground state energy for all
$J_z >0.02$. Of course numerically we cannot rule out that a phase transition
to a non uniform phase may actually occur for unphysically small $J_z$.
Anyway the fit of the energy data for small $J_z$ extrapolate
to the $J_z =0$ NK  energy with a remarkable accuracy (see caption Fig. 4b).

Thus the phase
separation  in the small $J$ $t-J$ model may be washed out by quantum
fluctuations\cite{rice,antimo}, especially in higher dimensions where
$E_{PS}$ approaches $e_F$ for $J_z \to 0$ with a much smaller  exponent.

We gratefully acknowledge useful discussions with E. Tosatti. This work
has been partially
supported by CNR under Progetto Finalizzato ``Sistemi informatici
e calcolo parallelo''.

\begin{figure}
\caption{Comparison of the LSD for 10, 18 and 26 (8,11 and 13)
 Lanczos iterations with corresponding analytical results
(solid lines)  within the  RP for the Bethe lattice with coordination number
$z=3$ ($z=4$).
Triangles, squares and circles correspond to the small,  medium
 and large $N$  calculation respectively.
The long dashed lines and dashed-dot lines are  fit of $A(\omega)$
obtained by the standard method  (see text) with
$\delta=0.05$ and $\delta=0.1$ respectively. }
\label{fig1}
\end{figure}
\begin{figure}
\caption{Calculated  $k=0$ -spectral weight and DOS for $N=10,18 ,26$
($8,11,13$)  for the 2C (2D) model.
 Solid lines are cubic interpolations
 of the largest $N$, and the
dashed lines are the BR densities  of states and a guide to the eye for the
2D-$A(k,\omega)$.
 The symbols for the points are as in Fig.1.}
\label{fig2}
\end{figure}
\begin{figure}
\caption{Calculated momentum dependent spectral function (lower part)
for different  $k$ in the
magnetic Brillouin zone and the dispersion of the first peak
for the same $k$ values (upper part). The bars are typical
estimates of the first-peak width.
For the 2C  $A(k,\omega)$, the wavevector $k$ ranges from
 from $(0,0)$ ( bottom) to $(\pi,0)$  (top) with nine equally spaced values.
For each $k$, we have shifted the spectral function by 0.25 successively.
For the 2D $A(k,\omega)$ the $k-$path in the magnetic Brillouin zone is shown
in the inset
The inset in the 2C dispersion is the free electron behavior.}
\label{fig3}
\end{figure}

\begin{figure}
\caption{(a):Plot of the lowest eigenvalues of the $2C$ model
as a function of $1/N$, the inverse of the  Lanczos iteration number,
for the same  momenta shown in Fig.3.
The horizontal dashed lines is  the BR ground state energy.
(b): ground state energies (crosses)  vs $J_z^{2/3}$ for the 2C model.
The continuous line connects linearly the  data and the dotted line is
an extrapolation ( consistent with the RP\protect\cite{vollhardt})   of the
last
five points for $J_z\to 0$:  $E(J_z) = E(0) +a {J_z\over t} ^{2/3} + b J_z$,
with $a/t= 2.6$, $b=-2.2$ and $E(0)=-2.996t$.
The long dashed line is the polaron energy (see text).}

\label{fig4}
\end{figure}

\end{document}